# Three-wave mixing experiments in indium-tin-oxide thin-films with no phase matching

Kyle Wynne[1], Marjan Bazian[2], and Mark C. Harrison[1]

[1]Fowler School of Engineering, Chapman University, Orange, USA
[2]Schmid College of Science and Technology, Chapman University, Orange, USA

E-mail: mharrison@chapman.edu

## Abstract

One challenge of using nonlinear optical phenomena for practical applications is the need to perform phase-matching. Recently, epsilon-near-zero materials have been shown to demonstrate strong optical nonlinearities, in addition to their other unique properties. As suggested by their name, the permittivity of the material is close to zero for a certain wavelength range. We demonstrate that this small permittivity allows for efficient three-wave mixing interactions to take place in indium-tin-oxide thin films without the need for phase matching the pump and signal beams. The efficiency of the second-order nonlinear interactions is characterized, and cascaded three-wave mixing is demonstrated.

Keywords: epsilon-near-zero, nonlinear optics, three-wave mixing, phase matching

## Introduction

Over the past few decades, remarkable scientific advancements have led to increased control over the properties of light. Among these advancements, nonlinear optics has played a significant role in enabling the manipulation of light in a variety of contexts. More recently, epsilon-near-zero (ENZ) materials have been recognized as useful nonlinear materials and the investigation of these materials has gained considerable attention (1). These materials have a permittivity ($\varepsilon$) that crosses zero at a specific wavelength (the zero-epsilon wavelength) and they exhibit many interesting properties in the wavelength range around this zero-crossing point. These unique properties make ENZ materials, such as indium tin oxide (ITO), useful for a variety of applications (2). For example, ITO's nonlinear properties can be used to tune its refractive index in useful ways (3). Furthermore, ENZ modes and surface plasmon polariton modes supported by these ENZ materials demonstrate substantial field enhancement within the material (4,5). Total absorption has been demonstrated, with applications in optoelectronics in the infrared at ambient temperatures (6–8). ENZ materials are not without their drawbacks. In particular, they can demonstrate high absorption in the ENZ region, which can be counterproductive when using them as a noninear material (9).

Many researchers have recently explored nonlinear optics in integrated photonics. It is often challenging to integrate nonlinear materials with conventional waveguide devices, forcing researchers to make tradeoffs to mitigate drawbacks (10). Unfortunately, the limitations inherent with on-chip nonlinear optics mean these methods are typically suitable for specific beam arrangements, leading to complex experimental setups or large devices. In particular, phase-matching is very difficult for integrated or waveguide-based nonlinear devices. However, by using low-index media like ENZ materials, we can satisfy phase-matching requirements for multiple input and output beam configurations, enabling more flexible and compact nonlinear devices. Due to the low index near the zero-epsilon wavelength (where nonlinearities arise), ENZ materials relax traditional phase-matching constraints, which can be exploited to facilitate nonlinear interactions and miniaturize nonlinear devices (3). Therefore, ENZ materials such as ITO are promising for use in integrated devices. We demonstrate three-wave mixing interactions in ITO thin films that require no phase matching and are a step towards using these materials in photonic integrated circuits.

**Background**

Manipulating and controlling light is a critical topic for photonics researchers (11,12). Because light interactions with matter are usually weak, making useful devices at the nanoscale is a significant challenge. Therefore, researchers have investigated many different macroscopic media to help control and manipulate light. For example, the use of engineered metamaterial leads to precise control of light, including bending, focusing, filtering, and even trapping and storing it (13). Researchers in this field have also studied ENZ materials for their attractive properties (3,4). The very low permittivity of these materials not only increases interactions between light and matter but also causes qualitatively different wave dynamics in the wavelength region close to the zero-permittivity wavelength. In addition, ENZ materials have demonstrated strong nonlinear optical interactions. These properties can be used for many applications such as compressing electromagnetic energy through very narrow channels, designing materials with a matching zero index, shaping the radiation pattern of a source, harmonic generation, and refractive index modulation (11,14,15). They can also be fabricated relatively easily, making them of interest to many researchers (16,17). Compared to conventional materials, ENZ materials have advantages such as fast response time, increasing the effect of slow light, and wide bandwidth (16).

Transparent conductive oxides like Al or Ga-doped zinc oxides and ITO are good candidates for nonlinear ENZ platforms (18–22). ITO is an ENZ material with large optical nonlinearities in the ENZ region around the zero-epsilon wavelength ($\lambda_{ZE}$). These materials support ENZ resonances in the near-infrared range and are compatible with existing complementary metal oxide semiconductor (CMOS) fabrication technologies, making them attractive for devices that could be fabricated at scale. The zero-epsilon wavelength of ITO films can be tuned in the near-infrared region from 1200 nm to 2900 nm via a simple annealing procedure, making it attractive for communications applications (23). Furthermore, transparent conductive oxides such as ITO and indium gallium-doped zinc oxide are used in industrial applications for touchscreen displays, transparent electrodes on solar panels, and more recently in televisions (24). Their commercial use makes them attractive for photonic devices because they are commercially available.

In this study, we focus on exploring second-order nonlinear effects in ITO, specifically three-wave mixing. Second order nonlinear effects are also known as $\chi^{(2)}$ nonlinearities. They are so named because $\chi^{(2)}$ is the second-order nonlinear coefficient for a material. The effects of $\chi^{(2)}$ nonlinearities include several three-wave mixing effects: second harmonic generation (SHG) or frequency doubling, sum frequency generation (SFG), and difference frequency generation (DFG).

In general, for three-wave mixing phenomena, the frequencies involved will obey the following relation:
$$\omega_1 + \omega_2 = \omega_3 \qquad (1)$$

In SHG, $\omega_1 = \omega_2 = \omega$ and $\omega_3 = 2\omega$. In this scenario, two photons of the same angular frequency ($\omega$) will combine in the nonlinear medium to one photon of double the wavelength (half the frequency). In SFG, $\omega_1$ and $\omega_2$ are used as the pump, and one photon from each frequency combines to produce a photon at $\omega_3$. In DFG, $\omega_3$ and $\omega_1$ are used as the pump and when they mix, they will produce a photon at $\omega_2$, as well as an additional idler photon at $\omega_1$.

The mixing relations given above do not capture all the dynamics of nonlinear optical interactions. When light is generated by three-wave mixing, the properties of the input beams dictate the propagation of the generated output light. The spatial distribution of the light beams within the material as well as the relative phase of the pump and generated beams have an impact, as seen in the equation below:
$$I_3 = \frac{1}{2}\left(\frac{\mu_0}{\varepsilon_0}\right)^{\frac{3}{2}} \frac{(\omega_3 \chi^{(2)} L)^2}{n_1 n_2 n_3} I_1 I_2 \left[\frac{\sin\left(\frac{\Delta k L}{2}\right)}{\frac{\Delta k L}{2}}\right]^2 \qquad (2)$$

This equation describes a plane-wave SFG interaction and is derived from a coupled wave equation analysis operating in the small signal limit. In this equation, the I terms are the intensity of light at each frequency, L is the interaction length, the n terms are the refractive index of the material at each frequency, $\varepsilon_0$ is the permittivity of free space, $\mu_0$ is the permeability of free space, and $\Delta k$ is the difference in k vectors between the $\omega_1$ and $\omega_2$ signals.

The $\Delta k$ dependence in equation (2) is required due to momentum conservation and is also known as the phase matching term. Typically, this term is made to be zero (or very small) in practical applications, a process which is called phase matching. Without phase matching, the generated intensity is severely limited. Some typical approaches to phase matching are quasi-phase matching, birefringent phase matching, and higher-order-mode phase matching. Nevertheless, using these approaches has some downsides. For example, they may only be phase-matched for specific arrangements of involved beams, often within a limited wavelength range. These phase matching methods all involve some external modification to the experimental setup, including adjusting angles of incident light or modifying materials. These constraints and modifications can place severe limitations on the compactness and flexibility of devices (25). They are a particular challenge for integrated photonic devices, where it may be difficult to phase match outside of bulk material processing (e.g. quasi-phase matching) or engineering specific modes and dispersion curves.

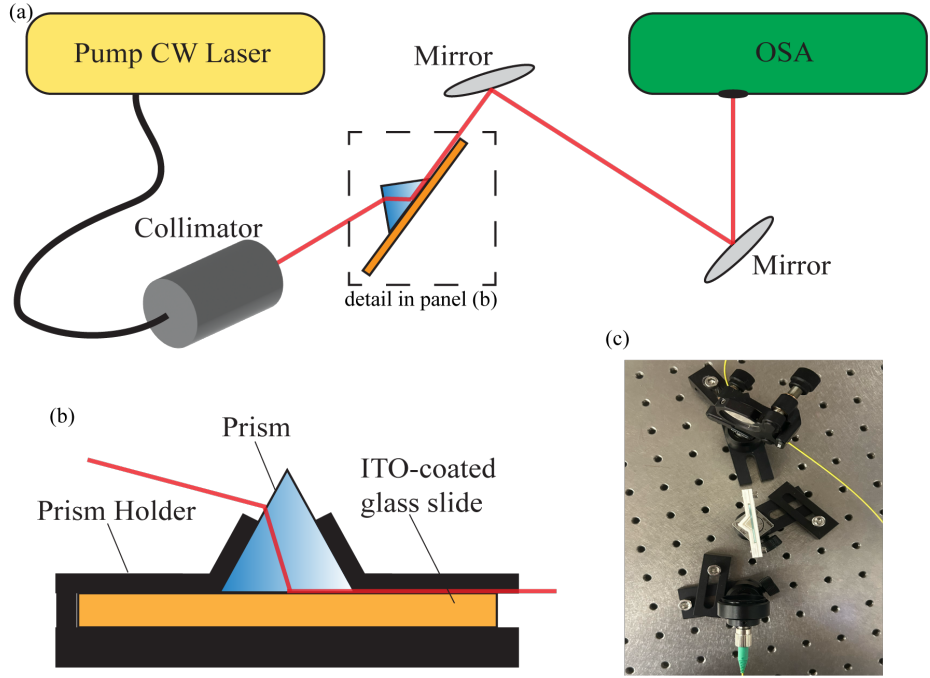

Figure 1. (a) Schematic of experimental setup. A laser is coupled into an ITO thin-film with a thickness between 120 – 160 nm using a collimator and prism. The output light is coupled into an optical spectrum analyzer (OSA) using mirrors. (b) Inset of the ITO slide and prism. The prism and slide are held in place by a custom 3D printed holder. (c) Photo of the setup showing the collimator, prism holder, and first mirror as seen from above.

ENZ materials also relax this phase matching term intrinsically. Because the light experiences such a low permittivity near the zero-epsilon wavelength, its k-vector will be very small, making $\Delta k$ close to zero. Another way to think of this is that the phase changes or accumulates extremely slowly for the wavelengths near the zero-epsilon wavelength, so will always be matched to other wavelengths (26,27). This lack of phase accumulation also relaxes the directional dependence of the phase matching requirement, which is different from most other nonlinear materials. This means that, in contrast to other nonlinear optical systems, ENZ materials reduce the need for external phase matching. In other words, the material itself helps facilitate phase matching intrinsically due to operating in a region where epsilon, and therefore the real part of the refractive index, is close to zero. Although this has been demonstrated over a wide wavelength range near the zero-epsilon wavelength of the ENZ material, this relaxed phase matching is distinct from broadband phase matching, in which phase matching is engineered for a broad wavelength range.

An important consideration for nonlinear optical interactions is the efficiency of power conversion from the pump beam to the generated beams. Typically, the efficiency ($\eta$) is given by:

$$\eta = \frac{I_{NL}}{I_P} \quad (3)$$

Where $I_{NL}$ is the intensity of the generated beam and $I_P$ is the intensity of the input, or pump, beam. Taking into account equation (2), we expect that the efficiency will depend on the $\chi^{(2)}$ coefficient, interaction length, and refractive index of the material. For small input power, the efficiency should vary roughly linearly with input intensity or input power. Because we expect the generated and fundamental beams to have roughly the same area in our experiments, in our efficiency calculations we use peak power instead of intensity. Using peak power is slightly less accurate, but results in a good estimate because the generated beam has a similar spectral linewidth to the fundamental beam.

**Methods**

We used ITO thin films with a thickness between 120 – 160 nm on glass slides. The ITO slides were purchased from Sigma Aldrich. The ITO thin film on a glass slide was paired with a coupling prism to help couple light into the thin film. The prism was held in the center of the slide, with the thin film facing the prism. By using the coupling prism in a modified Kretschmann configuration to couple light into the ITO thin film, we were able to extend the interaction length. Our measured interaction length is approximately 1.25 cm (half the width of the glass slide).

Later, to test the wavelength dependence of the nonlinear efficiency for different zero-epsilon wavelengths of the ITO material, we used an annealing process to adjust this wavelength. We used the same Sigma Aldrich ITO slides and the annealing process from *Johns, et. al.* (23). This process involved heating the samples to 350° C for varying lengths of time depending on the desired zero-epsilon wavelength. We followed the procedure to anneal different slides to have a target $\lambda_{ZE}$ of 1560 nm, 1460 nm, and 1350 nm. The non-annealed slide is predicted to have a $\lambda_{ZE}$ of 1280 nm. We also verified these zero-epsilon wavelengths for the annealed samples by using four-point probe measurements (see supplemental information). *Johns, et. al* report continuous and reversible tuning of the $\lambda_{ZE}$ from 1280 nm to 2900 nm (23).

*2.1 Basic Experimental Setup*

Our setup consisted of the following: a fiber coupled, CW laser attached to a collimator was used to couple light into an ITO thin film using a gadolinium gallium garnet prism (figure 1). The prism helped couple the light into a mode that propagates within the ITO film. The light coming out of the end of the thin film was aligned into a free-space port on an optical spectrum analyzer (OSA) from Thorlabs using two mirrors. We used a collimator designed for 1310 nm light. A real time spectrum was read from the OSA output which also gives the ability to track the center wavelength of peaks. Alignment was done by reading live power readings of the nonlinear signals measured by the OSA and adjusting the tilt position of the mirrors and the rotation of the prism holder to maximize the power. The prism holder was mounted on a post that allowed us to rotate the prism and slide configuration. The mirrors were set up using kinematic mirror mounts which allowed us to adjust the mirrors with 2 degrees of freedom each. With this setup we can align both the position of the beam and its angle entering the OSA. No phase matching was performed in the basic setup or any variation of this setup.

Our modified Kretschmann configuration was achieved using a 3D printed prism holder that could hold the prism in a fixed position relative to the ITO thin film (figure 1(b)). The slide was oriented so the ITO thin film faces the prism. Although the holder was designed to hold the prism in contact with the ITO slide, in practice there was a very small gap between the thin film and prism. Once the beam was aligned and sent through the prism holder, it was directed to a free-space port on the OSA.

All of our experiments were performed in a temperature-controlled laboratory, but we did not control the temperature of the sample beyond its ambient environment. Due to the relatively low power, CW lasers used, sample temperature and heating was not an issue. We confirmed this by tracking the conversion efficiency over two hours and saw no significant change (see supplemental information). Additionally, we did not control the polarization of the lasers and all light in our experiments was unpolarized. Since ITO is not a birefringent material and we are coupling into propagating modes in the thin film, we did not expect polarization to have an effect on the nonlinear interactions. We confirmed this by testing SFG with arbitrary polarization states and saw no significant change in efficiency (see supplemental information).

*2.2 Basic three-wave mixing*

There were several variations of this setup used to collect data for different experiments. The first variation was used to collect basic three-wave mixing data. First, we used a tunable CW diode laser with a range of 1270 – 1330 nm by itself as a pump laser and measured the generated second harmonic. In a slight variation we used a 2x2 fiber optic coupler to combine pump inputs from two different lasers. The first was the tunable CW diode laser set to 1330 nm and the second was a fixed wavelength CW diode laser at 1310 nm. The tunable lasers we used are of much higher quality than the fixed wavelength laser. The ITO slide used in this experiment was a non-annealed slide with a $\lambda_{ZE}$ of 1280 nm. The rest of the setup was unchanged from our basic setup.

*2.3 Zero-epsilon wavelength dependence*

The next set of experiments consisted of 4 complete sweeps through the supported wavelengths of two tunable CW lasers (1270 - 1330 nm, 1520 - 1570 nm) with a wavelength step interval of 5 nm. For each sweep we used one laser at a time and an ITO slide with a different $\lambda_{ZE}$. The four unique samples we used had target zero-epsilon wavelengths of 1280 nm, 1350 nm, 1460 nm, and 1560 nm. We tracked the peak power levels of the second harmonic while using the built-in power output monitoring functions of the tunable lasers to keep consistent pump power as we varied the wavelength. Both lasers were put at the minimum power level for generation of the beam which was 4.5 mW for the 1270 – 1330 nm and 12 mW for the 1520 – 1570 nm. Before data collection we verified that the power read for the pump beam was about even between the lasers with a slightly higher power read from the 1520 – 1570 nm laser. We recorded a spectra every 5 nm for each slide we had annealed. We continued to measure the spectra and tweak the alignment (which varies with wavelength) until we recorded a replicable power level for a given pump wavelength. We then recorded the power of the second harmonic. A power reading was not saved until it could be read multiple times for each datapoint.

*2.4 Power Vs. Efficiency*

Our next alternate setup was used to find the power conversion efficiency of the ITO slide with varying pump power. For this setup we used the tunable CW diode laser with a wavelength range of 1520 – 1570 nm. We used an ITO slide which was annealed to have a $\lambda_{ZE}$ of 1560 nm with the tunable laser set to a wavelength of 1550 nm. The power was first increased using the control panel on the laser to a maximum of 18.5 mW, and then increased using an erbium-doped fiber amplifier. The fiber amplifier was used to increase the power before sending the pump to the collimator. We then aligned the beam using the methods outlined in our basic setup and took data by tracking the peak power levels of the second

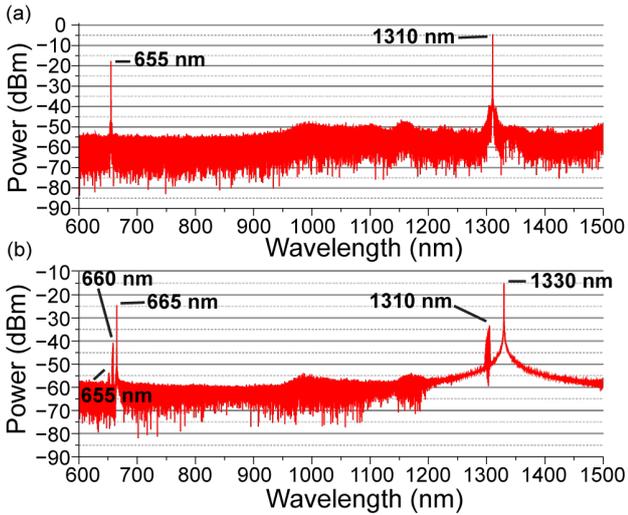

Figure 2. Three-wave mixing spectra generated by the ITO slide and measured by the OSA. (a) A tunable CW laser set to 1310 nm is used as a pump and the second-harmonic is visible at 655 nm. b) A tunable CW laser set to 1330 nm and a fixed-wavelength CW diode laser at 1310 nm are used as pumps. Their second harmonic signals at 655 nm and 665 nm are visible, along with a sum-frequency peak at 660 nm.

harmonic and the pump with software functions built into the OSA.

*2.5 Cascading nonlinear effects*

For the final experiment we used the 2x2 fiber optic coupler again. Instead of using a fixed wavelength laser, we used both of the high-quality tunable CW lasers. The tunable lasers can achieve a much higher peak intensity and have a narrower linewidth than the fixed wavelength diode laser from the basic three-wave mixing experiments. The high-quality CW lasers were mixed together at 4 different combinations of wavelengths. We used a single ITO sample annealed to a $\lambda_{ZE}$ of 1460 nm because that wavelength is between the wavelength ranges of the tunable CW lasers. This allowed for the most even power conversion efficiency levels between the two pump wavelengths.

**Results**

Our first experiments demonstrated basic three-wave mixing with a slide that had not been annealed. The results are shown in figure 2. We observed both SHG and SFG. For SHG in figure 2(a), we used a tunable 1310 nm laser and were able to generate a second harmonic at 655 nm with a power conversion efficiency of approximately 4.82%. The efficieny is calculated by using the power measured by the OSA. It would be better to measure the power input into the ITO, but we do not know the coupling efficiency of our prism configuration, so power measured by the OSA is a good estimate. Using a 2x2 optical fiber coupler and a fixed-wavelength diode laser as well as the tunable laser we generated a SFG spectra, shown in figure 2(b). The fixed-wavelength laser had a wavelength of 1310 nm and the tunable laser was set to 1330 nm. We observed peaks at 655, 660, and 665 nm corresponding to the second harmonic of the 1310 nm laser, the sum-frequency beam of the two lasers, and the second harmonic of the 1330 nm laser, respectively.

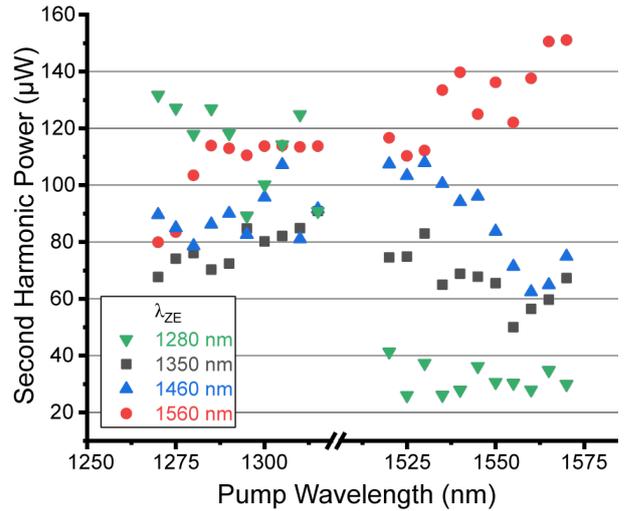

Figure 3. The second harmonic peak power vs. pump wavelength for slides annealed to have different zero-epsilon wavelengths. There is a gap in the x-axis between the wavelength ranges of our tunable CW lasers.

The next set of experiments involved sweeping the tunable lasers and measuring the SHG power for slides annealed to have different $\lambda_{ZE}$. The results are shown in figure 3 shape- and color-coded by the target $\lambda_{ZE}$ for the annealing recipe. We performed these experiments to quantify the behavior of the different slides at different wavelength ranges. The plot shows the second harmonic peak power recorded for each wavelength we measured. There is a trend that power conversion is more efficient the closer you are to the zero-epsilon wavelength of the ITO slide. We did not characterize the ENZ region for our samples explicitly, however the full wavelength range of our available tunable CW lasers produced measurable SHG peaks for all ITO slides. The gap in the center of the plot, indicated by a break in the x-axis, is due to the wavelength ranges available with the tunable lasers we used.

To examine how the conversion efficiency changes with respect to input power we compared the peak power of the second harmonic with the peak power of the fundamental as measured by the OSA. The results are shown in figure 4. We used a tunable laser set to 1550 nm. The conversion efficiency was calculated as the ratio of second harmonic peak power to pump peak power. Because we used the pump power as measured by the OSA (at the output of the thin film) instead of measuring the power that is coupled into the thin film (at the input), the efficiency might be slightly inflated. The trend for conversion efficiency in figure 4 is linear but eventually levels off. The initial linearly increasing region of the data on the left side of the chart and in the inset was taken by changing the power settings on the laser without using the fiber amplifier. At these lower powers, there was more noise in the measurement, resulting in the variation in the linear trend. At these powers, we expect there to be a linear increase in efficiency with increasing power, as we are probably operating within the small signal limit. However, once we start using a fiber amplifier, the second harmonic signal increases enough to start depleting the pump, resulting in a more constant efficiency. The slight downward trend in efficiency in this region is probably due to experimental error and an increase in loss. With the pump power increasing we may have started to lose extra power with small alignment issues. The lowest power measured is the lowest power that we could measure with the OSA, and the highest power level was the highest power we could measure with the OSA.

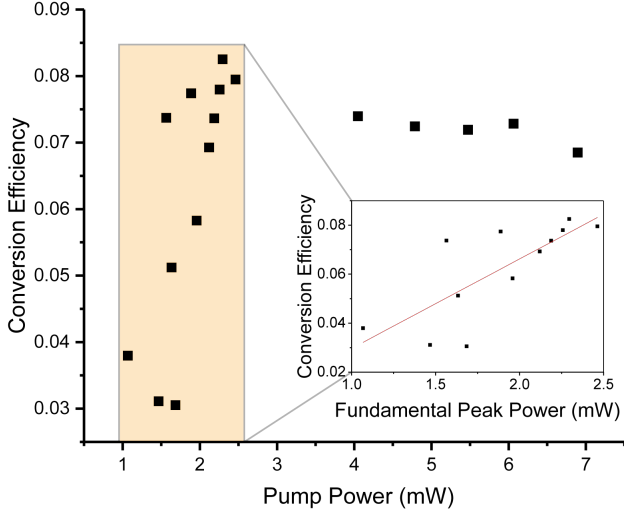

Figure 4. Conversion efficiency vs. measured pump power. The input wavelength was fixed at 1550 nm using a CW laser. After 18.5 mW input power (2.5 mW measured by the OSA), a fiber amplifier was used to increase the pump power. The inset shows the region before using a fiber amplifier with a linear fit to the data.

When performing wave mixing with both tunable CW lasers and a slide annealed to a zero-epsilon wavelength of 1460 nm, we observed cascaded wave-mixing effects in the nonlinear signal regeneration. We tested 4 different configurations of the two lasers. Each spectrum is recorded with the OSA, shown in figure 5. Table 1 shows which wave mixing phenomena and specific wavelengths could have generated each peak. The peak wavelengths in the table are the wavelengths measured by the OSA. The third harmonic falls outside of the range of our OSA with the cutoff being 600 nm. However, based on the observed mixing of other peaks, we were able to identify the third harmonic generation must be generated as well. The third harmonic generation is likely a cascaded three-wave mixing process as well, with the fundamental frequency mixing with the second-harmonic to produce the third-harmonic. These cascaded interactions display much higher efficiency than in figure 2 because both of the tunable lasers used in figure 5 are higher quality than the fixed wavelength laser. They also demonstrate the ability to have efficient nonlinear interactions involving wavelengths far away from the zero-epsilon wavelength of the ITO slide as long as one wavelength is close to the $\lambda_{ZE}$. We coupled the two lasers into the OSA directly without using an ITO slide and did not observe any peaks other than the laser peaks, confirming that these mixing peaks are not artifacts of the OSA.

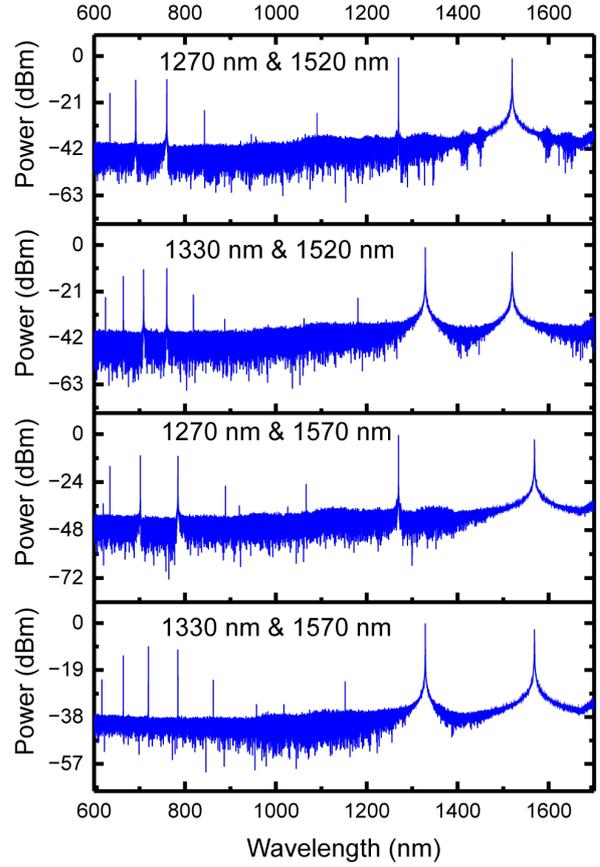

Figure 5. OSA spectra showing cascaded nonlinearities with two CW pump lasers using an ITO slide with a zero-epsilon wavelength of 1460 nm. The pump laser wavelengths for each spectrum are labeled at the top of each plot.

## Discussion

From figure 2(a) we were able to calculate efficiency a maximum efficiency of approximately 4.82%. While this is not the highest efficiency we measured, compared to conversion efficencies in other works, we can see that our efficiency compares favorably (28–30). This is likely due to a longer interaction length in our setup compared to previous works where the light travelled perpendicular to the ENZ thin-film plane. In contrast, due to our experimental setup, the pump light is likely coupled into an a long-range surface plasmon polariton mode or a Berreman mode that propagates in the thin film (3,5,30,31). This not only increases the interaction length, but likely increases the intensity because the guided mode has a smaller volume than the incident beam mode. The long interaction length also demonstrates that phase matching is not needed. The 1.25 cm interaction length is approximately 10,000x longer than the wavelength of light, so any large phase mismatch would severely limit nonlinear conversion efficiency.

Table 1
Peak wavelengths from figure 5 with wave-mixing descriptions

| 1330 1570 Mix Peaks | description | 1270 1570 Mix Peaks | description |
|---|---|---|---|
| 617.182 | THG + DFG: (1328 + 1328 + 1328) - 1568 | 620.352 | SFG: 1026 + 1568 |
| 664.395 | SHG: 1328 + 1328 | 634.921 | SHG: 1269 + 1269 |
| 719.429 | SFG: 1328 + 1568 | 701.806 | SFG: 1269 + 1568 |
| 784.405 | SHG: 1568 + 1568 | 784.443 | SHG: 1568 + 1568 |
| 862.282 | THG + DFG: (1568 + 1568 + 1568) - 1328 | 889.138 | THG + DFG: (1568 + 1568 + 1568) -1269 |
| 957.327 | THG + DFG: (1568 + 1568 + 1568) - 1152 | 919.363 | THG + DFG: (1269 + 1269 + 1269) - 784 |
| 1017.45 | THG + DFG: (1328 + 1328 + 1328) - 784 | 1026.07 | THG + DFG: (1568 + 1568 + 1568) -1066 |
| 1152.48 | SHG + DFG: (1328 + 1328) - 1568 | 1066.55 | SHG + DFG: (1269 + 1269) - 1568 |
| 1328.8 | pump | 1269.84 | pump |
| 1568.82 | pump | 1568.89 | pump |
| **1330 1520 Mix Peaks** | **description** | **1270 1520 Mix Peaks** | **description** |
| 625.116 | THG + DFG: (1328 + 1328 + 1328) - 1519 | 634.921 | SHG: 1269 + 1269 |
| 664.395 | SHG: 1328 + 1328 | 691.81 | SFG: 1269 + 1519 |
| 708.942 | SFG: 1328 + 1519 | 759.898 | SHG: 1519 + 1519 |
| 759.898 | SHG: 1519 + 1519 | 842.843 | THG + DFG: (1519 + 1519 + 1519) - 1269 |
| 818.738 | THG + DFG: (1519 + 1519 + 1519) - 1328 | 946.124 | THG + DFG: (1519 + 1519 + 1519) - 1090 |
| 887.456 | THG + DFG: (1519 + 1519 + 1519) - 1180 | 955.535 | THG + DFG: (1269 + 1269 + 1269) - 759 |
| 1061.89 | (THG + DFG) + DFG: 625 - 1519 | 1090.49 | SHG + DFG: (1269 + 1269) - 1519 |
| 1180.43 | SHG + DFG: (1328 + 1328) - 1519 | 1269.84 | pump |
| 1328.8 | pump | 1519.8 | pump |
| 1519.8 | pump | | |

The data in figure 2(b) further confirms efficient nonlinear interactions. In this set of experiments, we mixed two lasers, one of which was of poorer quality than the other. The poor-quality laser has a much lower peak power and broader spectral peak, and therefore lower intensity than the high-quality laser. The lower intensity causes a lower conversion efficiency for the poor-quality laser. Despite this lower intensity, we are able to observe not only wave mixing between the two sources (SFG), but also SHG from just the poor-quality laser by itself. We suspect that this would not be observable without the longer interaction length from light coupling into the ITO thin-film.

As seen in figure 3, the nonlinear efficiency has a clear dependence on the $\lambda_{ZE}$. The closer the pump wavelength is to the zero-epsilon wavelength, the higher the power of the second-harmonic wavelength signal, which is to be expected. The changing efficiencies are confirmation that we successfully shifted the zero-epsilon wavelength for these samples. While this has been demonstrated previously, as in *Capretti et al.*, in other works the different zero-epsilon wavelengths were a result of different ITO fabrication parameters (29). In the present work, we used commercial ITO that was fabricated using the same parameters, and adjusted the $\lambda_{ZE}$ using a separate procedure. Being able to adjust ITO to work at specific wavelengths is very important when considering using the material for integrated devices. It is likely that fabrication processes for waveguides and waveguide devices will shift the zero-epsilon wavelength away from the target wavelength. However, by annealing the samples, the zero-epsilon wavelength can be tuned to increase the efficiency of fabricated devices. This makes ITO a very flexible material that can be used in different circumstances.

When we investigated the power-dependence of the nonlinear interactions, as shown in figure 4, we see two distinct regions. In the first, the conversion efficiency varies linearly with input (pump) peak power. This is to be expected of nonlinear interactions, particularly those operating in the undepleted pump approximation, where we assume that the pump power is not being depleted. At lower input powers, this approximation is valid. However, when we start increasing the power further using the fiber amplifier, the efficiency levels off. In this regime, we believe that the pump is being depleted as the signal travels down the ITO until an equilibrium is reached. In other words, this efficiency is roughly the maximum efficiency for our experimental setup. The maximum efficiency we measure is actually 8.25%, but the equilibrium efficiency is approximately 7.5%. The efficiency is also related to the propagation length in the nonlinear material. Generally, a longer propagation length will result in a higher conversion efficiency until the pump is depleted up to some equilibrium. At that point, some of the second-harmonic signal could be converting back to the pump wavelength in a difference-frequency process. The second-harmonic could also mix back with the fundamental to create a third-harmonic signal, lowering efficiency. Unfortunately, the third-harmonic wavelength is outside the range of our OSA. Finally, a longer propagation length could also reduce efficiency if the material is lossy. Since ITO is an ENZ material and only the real part of the permittivity is close to zero, there is likely high loss for the propagating mode that the light couples into. This high loss could be a factor for the reduced efficiency we calculate in the equilibrium regime compared to the maximum from the linearly increasing region.

The cascaded nonlinear generation shown in figure 5 shows the potential of ITO as a nonlinear material. The different mixing of generated and pump wavelengths further demonstrates the utility of not needing any phase-matching for these experiments. Typically, this sort of cascaded generation requires a resonator or a very long interaction length (much longer than our 1.25 cm). The resonator would help phase match all the different wavelengths while providing a long interaction length. Here we see a large number of wavelengths, each with slightly different index, mixing and re-mixing with each other. The ability for all these wavelengths to mix in the sample indicates that no phase matching is necessary as long as one wavelength is close to the $\lambda_{ZE}$ of the material. Observing these cascaded peaks with a moderate interaction length and at relatively low input power bodes well for future applications involving ITO as a nonlinear material. The cascaded nonlinearities were achieved with a fairly simple experimental setup. Additionally, our measured conversion efficiency is lower than expected to be able to observe this cascaded interaction without some sort of resonator or high-powered pulses. By placing the material within a well-engineered resonator, we speculate that we could extend the proto-frequency comb shown in the figure into a full-fledged frequency comb, which could open up more application possibilities.

## Conclusion

In summary, we have demonstrated various three-wave mixing interactions inside of an ITO thin film using CW lasers and no phase matching in any of our experiments. We measured how the SHG efficiency changes with varying zero-epsilon wavelength of the material. We also characterized the power-dependence of the SHG efficiency, and saw that at a certain input power the efficiency will reach a maximum. Finally, we showed that under the right conditions we can observe cascaded nonlinear interactions which demonstrates the potential of using ITO as a nonlinear medium in photonic devices. Because these experiments involve a guided mode in a thin-film, they pave the way for using ITO as a nonlinear material for integrated devices, for which phase matching is a critical concern and a difficult problem to address.

## Acknowledgements

The authors would like to thank Nicholas Mirchandani who designed the initial version of the 3D printed slide-and-prism holder and did the preliminary setup for the experiments. This material is based upon work supported by the Air Force Office of Scientific Research (Award No. FA9550-21-1-0188). The authors have no conflicts of interest to disclose.

Supplemental Information

**4-point probe measurements**

In this study, a 4-point probe device is used to measure the resistance of samples. The samples we measured are ITO thin-films with a thickness between 120 nm and 160 nm. We use the 4-point probe measurements to measure the sheet resistance of the thin film. From the sheet resistance, we can estimate the zero-epsilon wavelength of the ITO. We measured samples with expected zero-epsilon wavelengths of 1280 nm, 1350 nm, 1460 nm and 1560 nm.

The 4-point probe setup includes a 3D-printed jig with two halves attached with small nuts and bolts as shown in supplementary figure 1. Each half has slots to hold 4 probe needles at equally spaced intervals of 2.41 mm. The needles are made of tungsten and have a 200 µm tip diameter. In order to perform a measurement, the needles are held in contact with the ITO thin film on the samples to measure the electrical resistance of the ITO. The 4 needles are adjusted to a constant height in the jig so that they make simultaneous contact with the sample surface while all the needle tips are in a straight line.

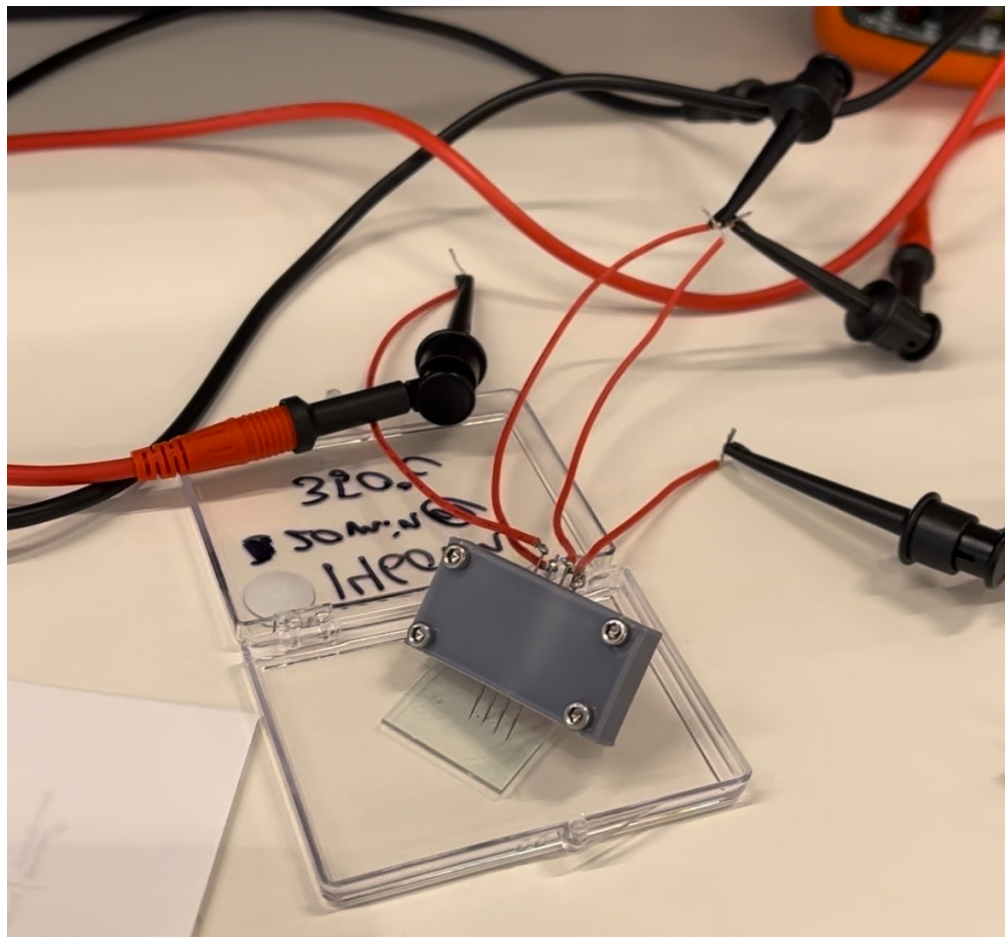

Supplementary figure 1. The custom 4-point probe jig on an ITO thin-film sample surface. The probe needles have leads soldered on the end to help connect them to the necessary power supply and voltmeter.

The outside probe needles are connected to a power supply set to 6 V with an ammeter connected in series. The middle two needles are connected to a voltmeter. When the probes are placed on the ITO thin film, the voltage between the inner needles and the current through the outer needles are measured. Using this measured voltage and current, the electrical resistance in ohms per square can be calculated according to the formula:

$$\rho_\square \left[\frac{\Omega}{square}\right] = \frac{\pi}{\ln(2)} \frac{V}{I}$$

We performed these measurements and the results summarized in supplementary table 1 below. We used these measurements compared with *Johns, et al.* to make sure our annealing procedure moved the zero-epsilon wavelength close to our target wavelength (1). Our 4-point probe results confirmed that our annealing procedure was successful, which is corroborated by the measurements shown in figure 3 in the main manuscript.

**Supplementary Table 1: Summary of sample properties for ITO films**

| sample | Annealing conditions | Target $\lambda_{ENZ}(nm)$ | Measured Resistance $\left(\frac{\Omega}{\square}\right)$ | Resistance in ref. (1) $\left(\frac{\Omega}{\square}\right)$ |
|---|---|---|---|---|
| S1 | untreated | 1280 | 10.23 | 12 |
| S2 | 350°C, 10 min, air | 1350 | 30.80 | 31.5 |
| S3 | 350°C, 20 min, air | 1460 | 31.1 | 36.6 |
| S4 | 350°C, 45 min, air | 1560 | 36.5 | 36.8 |

**Time-dependence measurements**

In this study we used CW lasers at relatively low input powers for most of our measurements, and we anticipated that our nonlinear experiments would be stable over time. In order to confirm this, we replicated our basic second-harmonic generation experiment using a 1310 nm laser and ENZ thin film. While monitoring the experiment over the course of two hours, we recorded a spectrum once every 15 minutes. We optimized the alignment before collecting the first spectrum, but did not adjust the alignment or any laser settings afterwards for the entire two hours. For each spectrum we recorded, we calculated the efficiency of the second-harmonic generation. The data from this experiment is shown in supplementary figure 2.

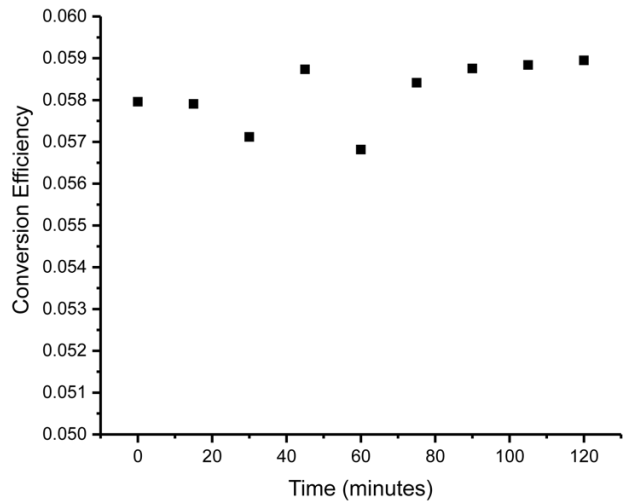

Supplementary Figure 2. Conversion efficiency vs. time. We recorded a spectrum once every fifteen minutes for a two-hour timespan and calculated the efficiency of the second-harmonic conversion. The efficiency remains stable over the course of two hours.

From the data, it is clear to see that the efficiency remains quite stable over the course of two hours. There are some small variations at two of the timesteps, and the efficiency increases slightly over time. The final efficiency is about 0.1 percentage points higher than the initial efficiency. However, these differences can easily be explained by noise in the measurement and the slow drift of the alignment mirrors. We did not use any temperature control in this experiment, but the laboratory where it was conducted is kept at a stable temperature. The data demonstrates that our experiment is stable over time. The ENZ thin film does not seem to heat up over time, nor does the nonlinear effect seem to saturate or diminish over time.

**Polarization-dependence measurements**

Typically, the polarization of pump beams is important in three-wave mixing in order to phase-match the signals involved in the nonlinear interaction. However, this is typically done in a birefringent crystal at a particular orientation to match the phase of all signals. ITO is not birefringent, so polarization should not help with phase matching as in a typical three-wave mixing experiment. If there is any polarization-dependence, it would be that the propagating mode in the thin film only supports a specific polarization. In order to confirm that three-wave mixing experiments are not polarization-dependent, we performed sum-frequency generation experiments with four different arbitrary polarization states for the pumps. In these experiments we used a 1310 nm laser and a 1550 nm laser as in the experiments described in section 2.5 in the main text. Before combining the lasers with the 2x2 fiber optic coupler, we used polarization controllers to set the polarization of each signal to an arbitrary state. We recorded data at four combinations of arbitrary states on each polarization controller, capturing ten spectra per polarization state. We calculated the efficiency as the peak power from the sum-frequency signal divided by the averaged peak power of the two pump signals. We calculated the mean and standard deviation for each polarization state. The data from this experiment is shown in supplementary figure 3.

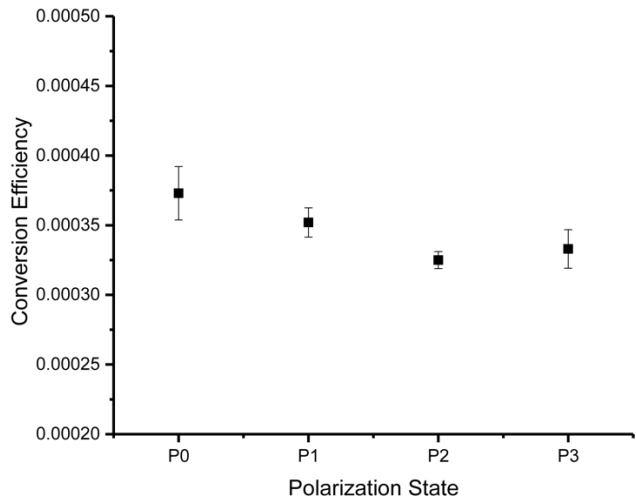

Supplementary Figure 3. Conversion efficiency vs. polarization state. We recorded ten spectra at four arbitrary polarization states of the two input pump beams. The sum-frequency generation efficiency was averaged and plotted with standard deviation used for the error. The efficiency remains stable over the course of two hours.

From the data, it is clear to see that the efficiency does not change much when the polarization state changes. In fact, we could not find any polarization state on either pump laser that prevented us from observing sum-frequency generation. This confirms not only that the nonlinear actions are not polarization-dependent, but also that external phase matching is not needed for our experimental setup. Polarizing the pump beams and using extra connections in the fiber optic link between lasers and the collimator introduced loss into the system, which resulted in lower pump powers and lower conversion efficiencies.